\documentclass{PoS}
\usepackage{amssymb,fontenc,times,mathptmx,graphicx}
\title{Distances to Galactic methanol masers}

\ShortTitle{Distances to Galactic methanol masers}

\author{\speaker{Kazi L.~J. Rygl}%
         \thanks{Member of the International Max Planck Research School for
           Astronomy and Astrophysics at the Universities of Bonn and Cologne.}\\
        Max-Planck-Institut f\"ur Radioastronomie, Germany \\
        E-mail: \email{kazi@mpifr-bonn.mpg.de}}
\author{Andreas Brunthaler\\
        Max-Planck-Institut f\"ur Radioastronomie, Germany \\
        E-mail: \email{brunthal@mpifr-bonn.mpg.de}}
\author{Karl M. Menten\\
        Max-Planck-Institut f\"ur Radioastronomie, Germany\\
        E-mail: \email{kmenten@mpifr-bonn.mpg.de}}
\author{Mark J. Reid\\
        Harvard-Smithsonian Center for Astrophysics, USA\\
        E-mail: \email{reid@cfa.harvard.edu}}
\author{Huib Jan van Langevelde\\
        Joint Institute for VLBI in Europe, the Netherlands\\
        Sterrewacht Leiden, Leiden University, the Netherlands\\
        E-mail: \email{langevelde@jive.nl}}

\abstract{We present the first EVN parallax measurements of 6.7
  GHz methanol masers in star forming regions of the Galaxy. The 6.7 GHz
  methanol maser transition is a very valuable astrometric tool, for its
  large stability and confined velocity spread, which makes it ideal to measure proper motions and parallaxes. Eight
  well-studied massive star forming regions have been observed during five EVN
  sessions of 24 hours duration each and we present here preliminary results for five of them. We achieve accuracies of up to 51 $\mu$as, which still have the potential to be improved by more ideal observational
  circumstances.}

\FullConference{The 9th European VLBI Network Symposium on The role of VLBI in the Golden Age for Radio Astronomy and EVN Users Meeting\\
		 September 23-26, 2008\\
		 Bologna, Italy}

\begin{document}

\section{Introduction}

In astronomy, accurate distances are a very important quantity to measure. 
First, the determination of many physical properties of objects, such as the size-scale,
the mass, the luminosity and the age depend on the distance. For
example, the
distance towards the Orion Nebula, determined by a trigonometric parallax of radio continuum
stars \cite{ment07}, turned out to be 10\% smaller than previously assumed. The
implications of this are a 10\% drop in the masses, 20\% fainter
luminosities and 20-30\% younger ages for the stars in the field. Moreover, one also needs distances on a large scale to study the structure of our Galaxy. 

Unfortunately, accurate distance measurements are very difficult. The only
method which allows an unbiased measure of distance is trigonometric parallax, which is simply a geometrical method, and therefore free of any
assumptions. To achieve the stability needed for optical wavelengths, parallax measurements require space-borne observations.
Historically, the first dedicated optical satellite for this
purpose was Hipparcos \cite{per95}, which had an accuracy of
0.8-2 mas, allowing it to measure distances to $\sim$200 pc (with error bars of
20\%). However, this successful experiment could reach only a tiny part of
the Milky Way. A new optical astrometry satellite GAIA, to be launched in 2012, will be able to
attain accuracies of 20 $\mu$as, and therefore provide important results on
Galactic-scale distances. Despite the high accuracy reachable in the
optical, GAIA will suffer from dust extinction, especially towards the
large and dense dust lanes in the Galactic plane. It is in these regions that
radio astronomy can provide an important complement to GAIA; radio
wavelengths do not suffer from dust extinction and VLBI phase referencing techniques
can provide accuracies up to 10~$\mu$as for obtaining accurate (with
error bars of 10\%) distances to 10~kpc (see for example
\cite{honma07},\cite{xu06} and \cite{reid08}). This is up to a 100-fold improvement on Hipparcos!

To measure the trigonometric parallax one requires very strong and compact
sources. Masers are the ideal targets, found frequently in (dusty) star
formation regions (SFRs) and AGB stars, objects whose distances one wants to know. Water masers at 22 GHz inhabit both
these regions, while methanol masers (12 and 6.7 GHz) are exclusively
associated with high mass SFRs. Water and 6.7 GHz methanol masers can be very strong,
with peak fluxes of the order of a kilojansky, but because of their stability and confined velocity spread methanol masers are preferred for astrometry. Water masers are often
associated with outflows in SFRs \cite{hachi06}, which
is manifested by a large velocity spread. Methanol masers are thought to be
pumped by emission from warm dust, heated by a young stellar
object and have velocity spreads of maximally $\sim$20 kms$^{-1}$, but
frequently much smaller \cite{sobo07}.
The frequency difference between water and methanol masers results in different
sources of disturbance for each of the transitions.
At higher frequencies the signal is more affected by tropospheric delay errors
than at lower frequencies, where the ionosphere is typically more dominant.
Also, the positional accuracy of the quasar one requires for phase referencing
depends on frequency. At high frequencies one observes the more optically
thick part of the quasar, while at lower frequencies one is dominated
by the optically thin part and hence one can encounter more extended structure.

When the project began in 2006, the EVN was the only VLBI network that had
receivers for observing the 6.7 GHz maser transition. 
It offered a unique opportunity to start a project measuring the parallaxes and
proper motions towards eight well known SFRs with strong methanol masers. Accurate
distances towards these targets will have an important impact on the study of
massive star formation in these regions. 

\section{Observations and data reduction}
The observations were performed at five epochs using the EVN between 2006, June
and 2008, March. The antennas participating in this
experiment were Jodrell Bank, Westerbork in single dish mode, Effelsberg, Medicina,
Torun, Onsala, Noto, Hartebeesthoek, Cambridge (first 3 epochs) and one VLA
antenna (last epoch). Each observation lasted 24 hours and made use of {\sl geodetic-like}
observations to calibrate the tropospheric delays. A typical observing run
contained 4 geodetic blocks of $\sim$1 hour every 6 hours, 4
blocks of fringe finders of $\sim$10 minutes, also every 6 hours, and 4 blocks of maser/quasar observations, spending on average between $\sim$0.9
and $\sim$1.2~hours on each maser. The difference in observing time per maser
depends on the sky position; half of our targets
are low declination sources, and have a limited visibility. 

Using the technique of phase-referencing, each maser was observed in a cycle
with one (or two) nearby ($\sim 1^\circ-2^\circ$) quasars. The observations were
performed with a data rate of 512 Mbps. The data were correlated at the Joint
Institute for VLBI in Europe (JIVE), using an integration time of 0.5 seconds
affording a field of view of 1.2 arc minutes (limited by time smearing).  
The maser data were correlated using one IF with high spectral resolution, with a bandwidth
of 8~MHz and a channel separation of 7.81 kHz, resulting in a spectral
resolution of 0.35 $\mathrm{kms}^{-1}$ at 6.7 GHz. The quasar sources were
correlated in continuum mode with 8 IFs, of 8 MHz bandwidth with a channel separation of 0.25 MHz. 

The data reduction was done using the Astronomical Image
Processing System (AIPS). The {\sl geodetic-like} observations were reduced
separately, and tropospheric delays were estimated for each antenna (see \cite{reid04}) and corrected with the task
'CLCOR'. The data were reduced following the EVN guidelines, applying
parallactic angle and ionospheric delay corrections. Amplitudes were
calibrated using system temperature measurements and standard gain curves. The
correction for the Earth rotation was done with the task 'CVEL'.
Since we observed strong masers we could use them as the
phase reference, and the solutions were transferred in a created 8IF dataset to
the continuum data. The quasar and maser positions were extracted by fitting 2D gaussians to the maps.
The differences in position between a maser and quasar pair were fitted, yielding the parallax and proper motions.
When more than one maser spot was used for the fit, the error was multiplied by $\sqrt N$, where $N$ is the number of
spots. This allows the random errors to be 100\% correlated, and is
therefore a conservative and safe approach.

\section{Preliminary results}

\begin{figure}[]
\centering
\includegraphics[width=6cm,angle=-90]{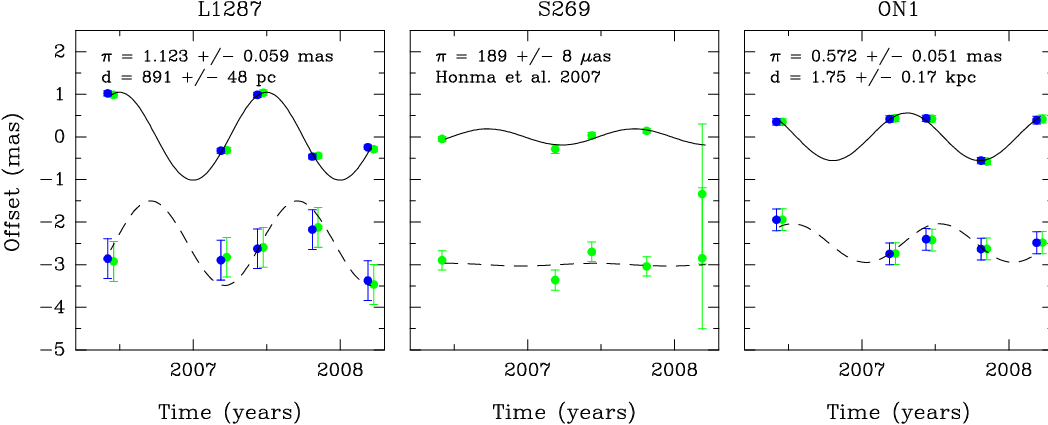}
\caption{The preliminary results for L1287, S269 and ON1. The Right Ascension
  (solid line) and declination (dashed line) signature of the parallax are
  shown. Each maser channel which was used for the parallax fitting is
  indicated by a different color.}
\end{figure}

\subsection{L1287}
L1287 is an interesting SFR, where CO observations \cite{yang91}
clearly indicate a bipolar outflow. An NH$_3$(1,1) study \cite{esta93} shows that there is cold and dense gas at the base of the outflow. In 2006, Imai et al.~found
evidence for a disk from the distribution of the radial velocities of water masers~\cite{imai06}.
We measure a parallax of $1.12\pm0.06$~mas (shown in Figure~1), corresponding
to a distance of $891\pm48$~pc -- very close to the kinematic
distance of 850 pc obtained from the ammonia spectral lines.

\subsection{S269}
In 2007, Honma et al.~measured the trigonometric parallax of
water masers to this SFR in the outer Galaxy to
be $189\pm8~\mu$as corresponding to a distance of $5.28\pm0.24$~kpc~\cite{honma07}. We were
not able to reach the necessary accuracy to verify the results of Honma et al.~with
our observations (Figure~1 shows the parallax of Honma et al. and our measurements). Nevertheless, we can study the
methanol proper motions and compare them to the proper motions obtained from
the water maser to test intrinsic differences between the methanol and water
maser distribution in this SFR.

\subsection{Overview}
A summary of the preliminary results is given in Table 1. Here, we briefly discuss the
remaining results. We measure the parallax of ON1 with an accuracy of
51~$\mu$as, corresponding to a distance of $1.75\pm0.17$~kpc (shown in Figure~1). This is the highest
accuracy we obtained due to a favorable sky position and a
good sampling of the parallax signature (peak, valley). For NGC281-W we
obtained the largest distance of $2.53\pm0.78$~kpc, which is in
excellent agreement with the water maser results \cite{sato08}. For S252 we did not manage to detect the quasar in
some epochs. However, S252 had a recent
12 GHz methanol maser parallax determined \cite{reid08}. The last three
sources (not in the Table), S255, L1206 and MonR2, are being worked on and we do not discuss them here.

\begin{table}[!h]
\centering
\caption{{\sc Preliminary distances and literature values}}
\begin{tabular}{llllll}
\\
\hline\hline
SFR & Parallax $\pm$ error & D$_{\mathrm{parallax}} \pm$ error & D$_{\mathrm{kinetic}}$ & Literature D\\
    & mas                 &kpc                     & kpc          & kpc         \\
\hline
L1287 & $1.123\pm0.059$  & $0.891\pm0.049$ & 0.850 & D$_{\mathrm{kin}}$; \cite{esta93}\\
ON1   & $0.572\pm0.051$ & $1.75\pm0.17$ & 1.8 & D$_{\mathrm{kin}}$; \cite{mac98}\\
NGC281-W & $0.395\pm0.1$ &$ 2.53\pm0.78$& 2.7 & 2.8; \cite{sato08}, parallax\\
S269  & - & - & 3.0 & 5.3; \cite{honma07}, parallax\\
S252 & - & - & 4.4 & 2.1; \cite{reid08}, parallax\\
\hline\hline
\end{tabular}
\end{table} 

\section{Discussion}

We have demonstrated that one can measure the trigonometric parallax of 6.7 GHz
methanol masers using the EVN. Good results were obtained for sources of
higher declinations ($\delta$~>~25$^\circ$). The accuracy reached in this
experiment is $\sim$50 -- 100~$\mu$as,
which is 10 -- 20 times better than Hipparcos. 
The accuracy of parallax measurements of the 6.7 GHz maser transition with the EVN can be improved by taking notice of the following. Our
project was large; it consisted of eight masers, all at different R.A.s and declinations. In
the five sessions, we did not manage to have the ideal parallax signature
coverage for each maser (compare L1287 with ON1 in Figure~1), for which one prefers to
sample the peaks. Often, as for L1287, some
sessions were near the zero-signature of the parallax, decreasing the quality of
the fit.
Another problem encountered in the project was the large number (50\%) of
low-declination sources. We did not manage to observe them as long as their
high declination counterparts. This, combined with their low elevations,
resulted in a poor $u,v$ coverage.

As in previous parallax studies, the position uncertainties in
declination are significantly worse than in Right Ascension (see for example
L1287 in Figure 1). First, the errors introduced by the atmosphere are larger in
declination. Second, the shape of the beam is more elongated in declination than in Right
Ascension. For example, the source S269 has a beam of $7.7\times3.7$~mas with a position angle of 55$^\circ$.

Finally, we would like to touch upon the {\sl geodetic-like} observations.
We find an improvement in the calibration for most of the cases.
However, in a few cases the corrections did not improve the calibration.
We suspect that here the ionospheric delay plays a more important
role, and actually we degrade the calibration by wrongly interpreting the determined
delay as of tropospheric origin. In these cases, we did not apply the 
{\sl geodetic-like} correction. A dual frequency observation mode, as employed in geodetic observations to calibrate the ionosphere, is not possible with most of the EVN
antennas. Also switching receivers is impossible. Therefore, determining the
S/N with or without a {\sl geodetic-like} correction in each observation is the best option for now.

In conclusion, we have shown that 6.7 methanol masers, observed with the EVN,
are a well suited tool for
astrometry. The fact that the VERA array was recently equipped with 5 GHz
receivers is a sign for a growing enthusiasm for 6.7 GHz methanol masers.
The final astrometry of our sources will be used to reevaluate the SFRs' astrophysics
in the light of the new distance determinations, and to study the 3D velocities of the SFRs relative to Galactic motion.

\end{document}